\documentclass[preprint,12pt]{elsarticle}

\usepackage{geometry}
\usepackage{amssymb}
\usepackage{amsmath}
\usepackage{amsfonts}
\usepackage{amssymb}
\usepackage[version=4]{mhchem}
\usepackage{stmaryrd}

\usepackage[colorlinks,
            linkcolor=blue,
            anchorcolor=blue,
            citecolor=blue]{hyperref}

\usepackage{booktabs}
\usepackage{multirow}
\usepackage{makecell}
\usepackage{threeparttable}
\usepackage{float}

\journal{}

\begin{document}

\begin{frontmatter}



\title{Viscous universe with cosmological constant }


\author[inst1]{Jinwen Hu}
\ead{200731890025@whu.edu.cn/2007.hujinwen@163.com}

\author[inst1]{Huan Hu}

\affiliation[inst1]{organization={Department of physics and technology},
            addressline={Wuhan university}, 
            city={Wuhan},
            postcode={430060}, 
            country={China}}

\begin{abstract}
We investigated a bulk viscous fluid universe with cosmological constant $\Lambda$ by assuming that the bulk viscosity to be proportional to the Hubble parameter. We found that for an expanding universe, the (relative) matter density will be always greater than a non-zero constant, and tends to this non-zero constant in the future. We show that the bulk viscosity model has a significantly better fitting to the combined $\mathrm{SNe} I a+\mathrm{CMB}+\mathrm{BAO}+H(z)$ data than the $\Lambda \mathrm{CDM}$ model. Generally, the evolution or values of some cosmological parameters predicted by the bulk viscosity model do not deviate significantly from which are obtained from the $\Lambda \mathrm{CDM}$ model since the bulk viscosity coefficient obtained from the astronomical observational data is so small. We also made a statefinder analysis of the bulk viscosity model and found that the evolution of the $\{r$, $s\}$ parameters behaves in such a way that $0<s<1,0.945<r<1$, indicating the bulk viscosity model is different from the $\Lambda \mathrm{CDM}$ model and the other ``dark energy" model.
\end{abstract}



\begin{keyword}
Cosmological constant \sep Bulk viscosity \sep Dark energy \sep $\Lambda \mathrm{CDM}$ model \sep Statefinder analysis

PACS: $95.36 .+\mathrm{x}$, 98.80.-k, \href{http://98.80.Es}{98.80.Es}
\end{keyword}
\end{frontmatter}


\section{Introduction}
Observations data, for instance the data from type Ia supernovae \cite{bib01,bib02}, the large-scale structure of universe \cite{bib03} and cosmic microwave background (CMB) \cite{bib04}, indicated that our universe is spatially flat and accelerating, which is considered that there exists an exotic cosmic fluid (called the dark energy) that accounts for about $2 / 3$ of the total energy of the universe. Despite the observational evidence on this existence of ``dark energy", its fundamental and nature origin is still unknown, resulting in many models have been proposed. Generally there are two different types of model. The first one is to modify the right side of the Einstein equation by introducing a special energy-momentum tensor $T_{\mu \nu}$ with a negative pressure. The simplest model for this type of model is the $\Lambda \mathrm{CDM}$ model, which introduce a cosmological constant $\Lambda$ as the dark energy, and the $\Lambda \mathrm{CDM}$ model is characterized by the equation of state, i.e. , $\omega_{\Lambda}=-1$. However, there exists some problems in the $\Lambda \mathrm{CDM}$ model, such as the coincidence problem \cite{bib05}, which refers to the coincidence that in today's universe the matter density and the dark energy density happen to be the same order of magnitude although they evolve at a different way. A possible answer to this question is that the dark energy is dynamical.

On the other hand, the process of dissipation arose in the viscous fluid, including both shear viscosity and bulk viscosity, as many studies shown, may be involved in the evolution of universe as an important role \cite{bib06,bib07,bib08}. Eckart \cite{bib09}, Landau and Lifshitz \cite{bib10} first studied the relativistic viscous fluid and derived the parabolic differential equations. But they just considered the first-order deviation from equilibrium, and from the obtained equations one can obtain that the heat flow and viscosity have an infinite speed of propagation, which is in contradiction with the principle of causality. Subsequently, Israel and Stewart \cite{bib11,bib12,bib13,bib14} studied the second-order deviation from equilibrium, and applied it to study the evolution of the early universe \cite{bib15}. However, in the context of the full causal theory the character of the evolution equation is so complicated that the simplified phenomenon, i.e., the quasi-stationary phenomenon, are studied by the conventional theory \cite{bib12,bib13,bib14}. In the isotropic and homogeneous universe, the dissipative process is usually characterized by a bulk viscosity parameter $\zeta$. And as the usual practice \cite{bib16}, the effect of the shear viscosity $\eta$ will be neglected. The dissipative effect caused by the bulk viscosity can be redefined by the effective pressure, $P_{\text {visc }}=-3 \zeta H$, where $\zeta$ is the bulk viscosity coefficient and $H$ is the Hubble parameter. And with the requirement of the second law of thermodynamics, it requires that $\zeta>0$, which assures that the entropy production is positive \cite{bib17,bib18}.

In the late accelerating universe, the effect of bulk viscous fluid was studied in Refs. \cite{bib19,bib20,bib21,bib22,bib23}. But a drawback faced by the bulk viscous fluid is to find a valid mechanism for the origin of bulk viscosity in the expanding universe. According to some theoretical point of view, when the local thermodynamics equilibrium is broken, the bulk viscosity arises \cite{bib24}. And the bulk viscosity can be regarded as an effective pressure to keep the system back to its thermal equilibrium state. The bulk viscosity pressure arises when the cosmological fluid expands or contract too fast (i.e., the state deviated from the local thermodynamics equilibrium) \cite{bib25,bib26,bib27}, and ceases when the fluid reaches the thermal equilibrium again.

In an expanding universe, the expansion process is likely to be a collection of states that lose their thermal equilibrium in a small fraction of time \cite{bib28}. It is therefore natural to assume that there exists a bulk viscosity coefficient in the description of the universe. In this paper, we investigated a bulk viscous matter universe with cosmological constant $\Lambda$, presenting a model with equivalent dynamical ``dark energy". This paper is organized as follows. In Sect. \ref{sec2} we present the basic formalism of the flat universe with bulk viscous fluid (called the bulk viscosity model) and show the interesting conclusions from this model. And we derived the evolution equation of the Hubble parameter. In Sect. \ref{sec3}, we investigated the $\Lambda \mathrm{CDM}$ model and the bulk viscosity model with the observational data from SNe Ia, CMB, BAO and Hubble parameter. In Sect. \ref{sec4} we present the statefinder analysis of the bulk viscosity model. In Sect. \ref{sec5} we summarized this paper.

\section{FLRW universe with bulk viscous matter}\label{sec2}
We consider a spatially flat universe described by the Friedmann-Lemaitre-RobertsonWalker (FLRW) metric \cite{bib29,bib30}
\begin{equation}
    \label{eq1}
    d s^{2}=-d t^{2}+a(t)^{2}\left(d r^{2}+r^{2} d \theta^{2}+r^{2} \sin ^{2} \theta d \phi^{2}\right)
\end{equation}
where $(r, \theta, \phi)$ are defined as the co-moving coordinates, $t$ denotes the proper time of cosmic, and $a(t)$ denotes the scale factor of the universe with bulk viscous matter, which produce an equivalent pressure \cite{bib10,bib31}
\begin{equation}
    \label{eq2}
    P_{e f f}=P-3 \zeta H
\end{equation}
where $H$ is the Hubble parameter, $P$ denotes the normal pressure, which is equal to 0 for the nonrelativistic matter, and $\zeta$ is generated in the viscous fluid that deviates from the local thermal equilibrium, and $\zeta$ can be a function of Hubble parameter and its derivatives. Here we didn't consider the radiation component for it is a reasonable simplification to the late universe.

Based on Eq. (\ref{eq2}), for an imperfect fluid, and with the first-order deviation from the thermodynamic equilibrium, we can obtain the form of the energy-momentum tensor \cite{bib32}
\begin{equation}
    \label{eq3}
    T_{\mu v}=\rho u_{\mu} u_{v}+\left(g_{\mu v}+u_{\mu} u_{v}\right) P_{\text {eff }}
\end{equation}
where $\rho$ denotes the density of the fluid.

It can be seen that the above energy-momentum tensor is similar to that of a perfect but with an equivalent pressure $P_{\text {eff }}$, which is composed by the pressure $P$ of the fluid plus the viscous pressure $P_{\text {visc }}=-3 \zeta H$. And as stated above that the viscous pressure, $P_{\text {visc }}$, can be regarded as a ``measurement" of the pressure to restore the local thermodynamic equilibrium.

Taking the cosmological constant $\Lambda$ into consideration, we conclude that for a pressureless matter universe $(P=0)$ governed by General Relativity in the presence of a viscous fluid, and with the FLRW geometry the two Friedman equations can be written as
\begin{equation}
    \label{eq4}
H^{2}=\frac{8 \pi G}{3} \rho_{m}+\frac{\Lambda}{3} 
\end{equation}

\begin{equation}
    \label{eq5}
\frac{\ddot{a}}{a}=-\frac{4 \pi G}{3}\left(\rho_{m}-9 H \zeta\right)+\frac{\Lambda}{3}
\end{equation}

Where $\rho_{m}$ is the total matter density, $G$ is the gravitational constant, $\Lambda$ denotes the cosmological constant, and the dot means time derivative. The conservation equation corresponding to Eq. (\ref{eq3}) is
\begin{equation}
    \label{eq6}
    \dot{\rho}_{m}+3 H\left(\rho_{m}-3 H \zeta\right)=0
\end{equation}

For an expanding universe (i.e., $H>0$ ), we have
\begin{equation}
    \label{eq7}
    \dot{\rho}_{m}<0
\end{equation}

Therefor, based on Eq. (\ref{eq6}), we obtain
\begin{equation}
    \label{eq8}
    \rho_{m}>3 H \zeta
\end{equation}

Substituting Eq. (\ref{eq4}) into Eq. (\ref{eq8}), we obtain
\begin{equation}
    \label{eq9}
H^{2}-8 \pi G \zeta H-\frac{\Lambda}{3}>0
\end{equation}

Note that Eq. (\ref{eq9}) always holds in an expanding universe. Here we defined the dimensionless Hubble parameter $E$, the present (relative) vacuum density parameter $\Omega_{\Lambda 0}$ and the present (relative) matter density parameter $\Omega_{\mathrm{m} 0}$ as
\begin{equation}
    \label{eq10}
E=\frac{H}{H_{0}}, \quad \Omega_{\Lambda 0}=\frac{\Lambda}{3 H_{0}^{2}} \quad \Omega_{m 0}=\frac{8 \pi G}{3 H_{0}^{2}} \rho_{m 0}
\end{equation}

Where $H_{0}$ is the present value of the Hubble parameter, $\rho_{m 0}$ is the present matter density.

On the other hand, based on Eqs. (\ref{eq4}) and (\ref{eq5}), we have
\begin{equation}
    \label{eq11}
\dot{H}=-4 \pi G\left(\rho_{m}-3 H \zeta\right)
\end{equation}

Obviously, it can be seen that the evolution of $H$ is related to the parameter $\zeta$. But so far we still don't know the explicit form of the bulk viscosity, which has to be usually assumed a priori (see for instance Refs. \cite{bib33,bib34,bib35,bib36,bib37,bib38}, the bulk viscosity coefficient is assumed a prior without being derived from known physics). In Refs. \cite{bib39,bib40,bib41,bib42,bib43,bib44,bib45,bib46,bib47,bib48}, it is assumed that the parameter $\zeta$ is relate to the Hubble parameter, i.e., $\zeta=\zeta(H)$, and one of the assumed forms is
\begin{equation}
    \label{eq12}
\zeta=\zeta_{0}+\zeta_{1} H
\end{equation}

Where $\zeta_{0}, \zeta_{1}$ are constants. The motivation for considering this form of bulk viscosity comes from the fluid mechanics, which holds that the viscosity phenomenon is related to the ``velocity". And both $\zeta=\zeta_{0}$ and $\zeta=\zeta_{1} H$ are separately studied by many authors.

Here we take the simple case of $\zeta=\zeta_{1} H$ as the bulk viscosity model, which follows the assumption on the parameter $\zeta$ in Refs. \cite{bib39,bib43,bib45,bib48,bib49}.

The other reason why we take the simple case of $\zeta=\zeta_{1} H$ is that, substituting $\zeta=\zeta_{1} H$ into Eq. (\ref{eq9}), we can obtain
\begin{equation}
    \label{eq13}
\Omega_{m}>8 \pi G \zeta_{1}
\end{equation}

Eq. (\ref{eq13}) indicates that for an expanding universe, the (relative) matter density $\Omega_{\mathrm{m}}$ will be always greater than a non-zero constant, i.e., $8 \pi \mathrm{G} \zeta_{1}$, and tends to this non-zero constant in the future, which is an interesting model, and based on Eq. (\ref{eq11}), it indicates that the universe in the bulk viscosity model tend to the de Sitter time-space in the future, which is similar to the $\Lambda \mathrm{CDM}$ model, but the expansion rate of the future's universe predicted by the bulk viscosity model is lower than which predicted by the $\Lambda \mathrm{CDM}$ model.

Based on Eqs. (\ref{eq4}), (\ref{eq5}) and (\ref{eq10}), we can obtain
\begin{equation}
    \label{eq14}
\frac{d E}{d z}=\frac{\left(1.5-\Omega_{\tau}\right) E}{1+z}-\frac{1.5 \Omega_{\Lambda 0}}{E(1+z)}
\end{equation}

Where $\Omega_{\tau}=12 \pi G \zeta_{1}$ is a dimensionless parameter, $z$ denotes the cosmological red shift, and $a=1 /(1+z)$.

\section{Parameters estimation}\label{sec3}
\subsection{Model constraints from SNe Ia alone}\label{sec3.1}
\begin{figure}
    \centering
    \includegraphics[width=0.6\linewidth]{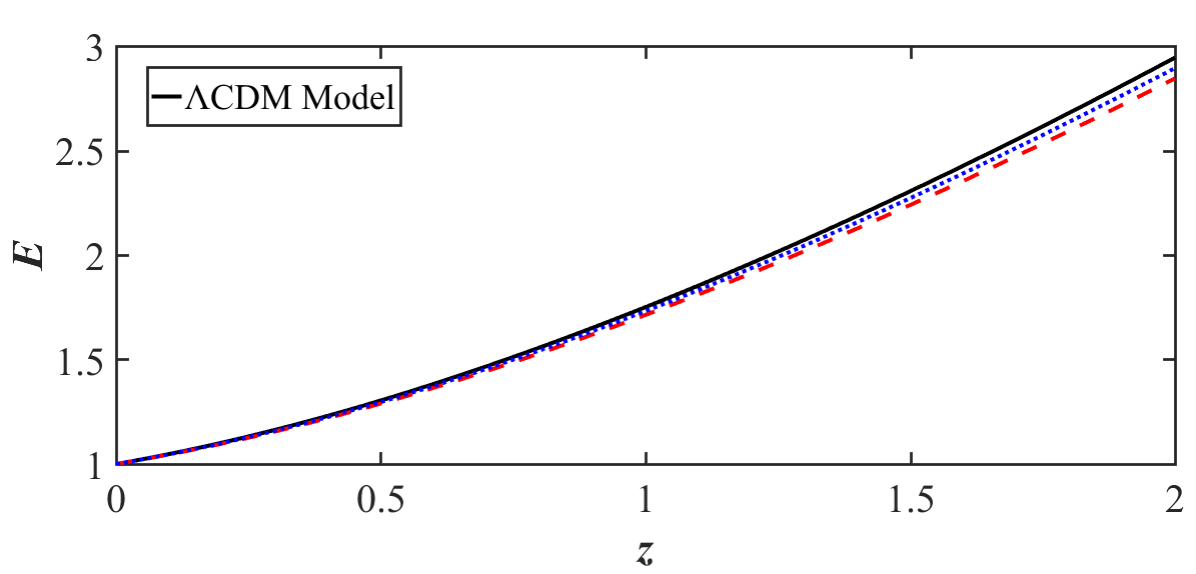}
    \caption{Behavior of the parameter $E$ when taking $\Omega_{\mathrm{m} 0}=\mathbf{0 . 2 9 5}, \Omega_{\Lambda 0}=0.705$ for the two models. The solid line corresponds to the $\Lambda$ CDM model, the dotted line corresponds to the bulk viscosity model with $\Omega \tau=0.01$, the dashed line corresponds to the bulk viscosity model with $\Omega \tau=\mathbf{0 . 0 2}$.}
    \label{fig01}
\end{figure}

It is well known that the $\Lambda \mathrm{CDM}$ model has been constrained by the observational data, such as the SNe Ia data, Power spectrum of the Cosmic Microwave Background fluctuations (CMB), Baryon Acoustic Oscillations (BAO), etc. Now we first consider the SNe Ia data alone. In Ref. \cite{bib50}, with the JLA SNe Ia sample, the authors found that the best fit value for $\Omega_{\mathrm{m} 0}$ is $0.295 \pm 0.034$ with a fixed fiducial value of $H_{0}=70 \mathrm{~km} / \mathrm{s} / \mathrm{Mpc}$. So here we first take $\Omega_{\mathrm{m} 0}=0.295, \Omega_{\Lambda 0}=1$ $\Omega_{\mathrm{m} 0}=0.705$ in Eq. (\ref{eq14}) to study the evolution of the parameter $E$, which is shown in figure \ref{fig01}.

In a spatially flat universe, the luminosity distance $d_{\mathrm{L}}$ is defined as \cite{bib50}
\begin{equation}
    \label{eq15}
d_{L}=(1+z) \int_{0}^{z} \frac{d z^{\prime}}{H}
\end{equation}

And the theoretical distance module $\mu_{t}$ for the $k$ th Supernova at red shift $z_{k}$ is defined as
\begin{equation}
    \label{eq16}
\mu_{t}=m-M=5 \log _{10}\left(\frac{d_{L}}{\mathrm{Mpc}}\right)+25
\end{equation}

Where $m$ and $M$ are the apparent and absolute magnitudes of the $\mathrm{SNe}$, respectively.

\begin{figure}
    \centering
    \includegraphics[width=0.6\linewidth]{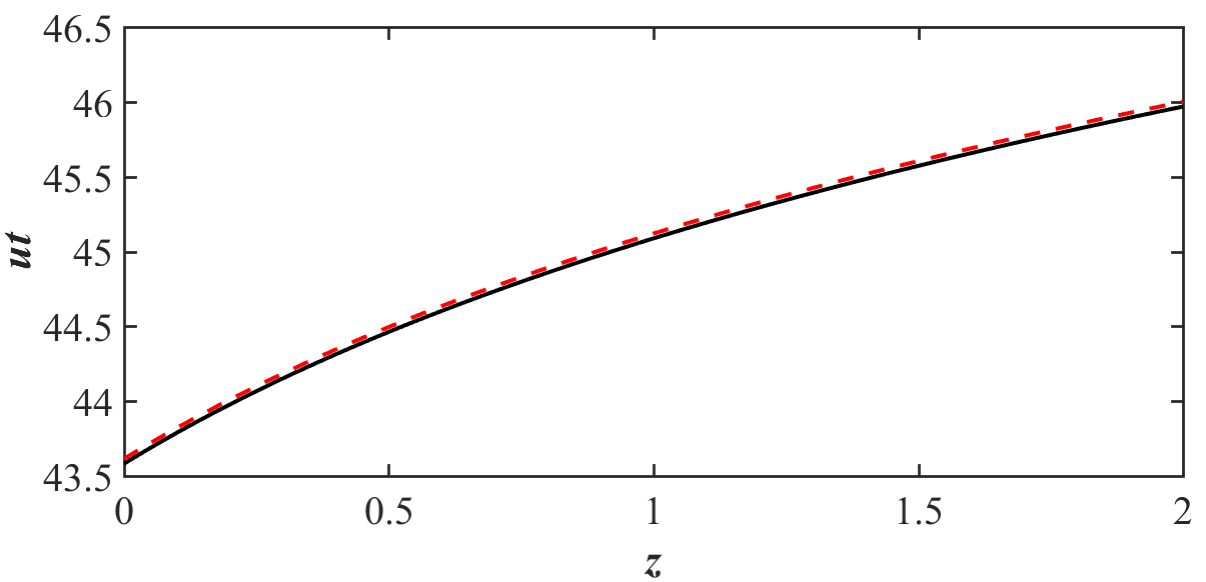}
    \caption{The comparison of the $\mathrm{z} \sim \mu_{t}$ curve when taking $\Omega_{\mathrm{m} 0}=\mathbf{0 . 2 9 5}, \Omega_{\Lambda 0}=0.705, H_{0}=70$ $\mathrm{km} / \mathrm{s} / \mathrm{Mpc}$ for the two models. The solid line corresponds to the $\Lambda \mathrm{CDM}$ model, the dashed line corresponds to the bulk viscosity model with $\Omega \tau=0.02$.}
    \label{fig02}
\end{figure}

Figure \ref{fig02} shows the comparison of the $\mathrm{z} \sim \mu_{t}$ curves derived from the $\Lambda \mathrm{CDM}$ model and bulk viscosity model with the same fiducial value of $\Omega_{\mathrm{m} 0}=0.295, \Omega_{\Lambda 0}=1-\Omega_{\mathrm{m} 0}=0.705, H_{0}=70 \mathrm{~km} / \mathrm{s} / \mathrm{Mpc}$.

Note that the JLA SNe Ia sample adopted in this paper cover the red shift range $0.01<z<1.2$. And from Figure \ref{fig01} and Figure \ref{fig02} it can be seen that the two curves are basically coincident or have little deviation $(<1 \%)$ from each other within the range $0.01<z<1.2$. With the negligible deviation $(<1 \%)$ of the parameter $E$ and $\mu_{t}$, we can conclude that within the red shift range $0.01<z<1.2$, the best estimated parameters corresponding to the two models are almost the same. That is, for the bulk viscosity model proposed in this paper, with a fixed fiducial value of $H_{0}=70 \mathrm{~km} / \mathrm{s} / \mathrm{Mpc}$, the best estimated parameters obtained from the JLA SNe Ia sample are $\Omega_{\mathrm{m} 0}=0.295 \pm 0.034$, $\Omega_{\Lambda 0}=0.705 \pm 0.034, \Omega_{\tau}=0.01 \pm 0.01$.

\subsection{Data from Power spectrum of the Cosmic Microwave Background fluctuations}
The temperature power spectrum of CMB is sensitive to the physics of decoupling epoch and the physics between today and the decoupling epoch. The former mainly affects the ratio of the peak heights and the Silk damping, i.e., the amplitude of acoustic peaks, while the latter affects the locations of peaks via the angular diameter distance out to the decoupling epoch. The ``acoustic scale" $l_{A}$ and the ``shift parameter" $R$ is defined as \cite{bib51}
\begin{equation}
    \label{eq17}
l_{A}=\frac{\pi r\left(z^{*}\right)}{r_{s}\left(z^{*}\right)}, \quad R=\sqrt{\Omega_{m 0}} \int_{0}^{z^{*}} \frac{d z^{\prime}}{E\left(z^{\prime}\right)}
\end{equation}

Where $z^{*}$ is the red shift at decoupling, $r(z)$ is the co-moving distance, which is defined as
\begin{equation}
    \label{eq18}
r(z)=\frac{1}{H_{0}} \int_{0}^{z} \frac{d z^{\prime}}{E\left(z^{\prime}\right)}
\end{equation}
and $r_{\mathrm{s}}\left(\mathrm{z}^{*}\right)$ is the co-moving sound horizon at recombination, which is defined by
\begin{equation}
    \label{eq19}
r_{s}\left(z^{*}\right)=\int_{0}^{a\left(z^{*}\right)} \frac{c_{s}(a)}{a^{2} H(a)} d a
\end{equation}

Where the sound speed $c_{\mathrm{s}}(a)$ is defined by
\begin{equation}
    \label{eq20}
c_{s}(a)=\left[3\left(1+\frac{3 \Omega_{b 0}}{4 \Omega_{\gamma 0}}\right)\right]^{-1 / 2}
\end{equation}

To determine the optimal value of the parameters (with $1 \sigma$ error at least) involved in the dark energy model, we will adopt the maximum likelihood method. That is, when the observation data satisfy the Gaussian distribution, then the corresponding likelihood function will satisfy $L \propto e^{-x^{2} / 2}$, where the statistical function $\chi^{2}$ represents the deviation between the theoretical value derived from a certain model and the observed value. The maximum likelihood method refer to finding the optimal value of the parameter involved in the theoretical model to minimize the value of $\chi^{2}$, and when the observation data we use to restrict the theoretical model are not independent of each other, such as the above three observation data from $\mathrm{CMB}$, i.e., $l_{A}, R, z^{*}$, then the $\chi^{2}$ function corresponding to the CMB data can be given as \cite{bib52}
\begin{equation}
    \label{eq21}
\chi_{C M B}^{2}=X^{T} C_{C M B}^{-1} X
\end{equation}

Where
\begin{equation}
    \label{eq22}
X=\left(\begin{array}{l}
l_{A}-302.40 \\
R-1.7246 \\
z^{*}-1090.88
\end{array}\right)
\end{equation}
and the inverse covariance matrix is given by \cite{bib52}
\begin{equation}
    \label{eq23}
C_{C M B}^{-1}=\left(\begin{array}{llc}
3.182 & 18.253 & -1.429 \\
18.253 & 11887.879 & -193.808 \\
-1.429 & -193.808 & 4.556
\end{array}\right)
\end{equation}

Based on Eqs. (\ref{eq17}) $\sim$ (\ref{eq23}), and with the best estimated parameters obtained from the JLA SNe Ia sample above, we can calculate the $\chi^{2}$, which is shown in Table \ref{tab01}.

\begin{table}[h]
    \centering
    \caption{Computation of the parameters in $\mathrm{CMB}^{\mathrm{a})}$}\label{tab01}
    \begin{tabular}{ccccccc}
\toprule

\multirow{3}{*}{Parameters $^{\mathrm{b})}$ }& \multicolumn{2}{c}{ \makecell[c]{Theoretical value\\ from models}} & \multirow{3}{*}{\makecell[c]{Observed\\mean value\\in Ref. [42] }  } & \multicolumn{2}{c}{$\chi^{2}$} & $\chi^{2} \min$ \\
\cmidrule{2-3} \cmidrule(r){5-6} \cmidrule(r){7-7}

&\makecell[c]{ $\Lambda \mathrm{CDM}$ \\ model }&\makecell[c]{ Bulk viscosity\\ model} &  & \makecell[c]{ $\Lambda \mathrm{CDM}$ \\ model } &\makecell[c]{ Bulk viscosity\\ model}& \makecell[c]{ $\Lambda \mathrm{CDM}$ \\ model }\\
\midrule

$z^{*}$ & 1090.88 & 1090.88 & 1090.88 &  &  &  \\

$l_{\mathrm{A}}$ & 337.22 & 302.56 & 302.40 & 3887.78 & 13.75 & 1280.06 \\

$R$ & 1.7468 & 1.8293 & 1.7246 &  &  &  \\

\bottomrule
    \end{tabular}
     \begin{tablenotes}
        \footnotesize
 \item  a) Here we adopt the best estimated parameters obtained from the JLA SNe Ia sample in Sect. \ref{sec3.1}, i.e., in the $\Lambda$ CDM model $\Omega_{\mathrm{m} 0}=0.295, \Omega_{\Lambda 0}=0.705$ and in the bulk viscosity model $\Omega_{\mathrm{m} 0}=0.295$, $\Omega_{\Lambda 0}=0.705, \Omega_{\tau}=0.018$ (within the range of $\Omega_{\tau}=0.01 \pm 0.01$ ), instead of computing the ``best estimated" for parameters from the CMB sample alone, for the purpose to directly compare the two models.

 \item b)The computation method of $z^{*}$ can be seen in Ref. \cite{bib53}.
      \end{tablenotes}

\end{table}

Table \ref{tab01} shows the value of $\chi^{2}$ obtained from the bulk viscosity model is significantly less than the value of $\chi^{2}$ and $\chi^{2}$ min $\left(\chi^{2}\right.$ min corresponds to the $\Lambda \mathrm{CDM}$ model with the best estimated parameters obtained from the CMB data alone) obtained from the $\Lambda \mathrm{CDM}$ model, which imply the bulk viscosity has a significantly better fitting to the combined JLA SNe Ia and CMB data than the $\Lambda \mathrm{CDM}$ model.

\subsection{Data from Baryon Acoustic Oscillations}
In Ref. \cite{bib52}, it shows that the acoustic peak in the galaxy correlation function has been detected in the red shift range $z=0.1 \sim 0.7$. This linear feature in the galaxy data provide a standard ruler with which to measure the distance ratio, $r_{\mathrm{s}} / D_{v}$. And the position of the $\mathrm{BAO}$ can be used to constrain $d_{z} \equiv r_{\mathrm{s}}\left(z_{d}\right) / D_{v}(z)$, where $r_{s}\left(z_{d}\right)$ is the co-moving sound horizon at the baryon drag epoch, and the $D_{v}(z)$ is defined by \cite{bib54}
\begin{equation}
    \label{eq24}
D_{v}(z) \equiv\left[\left(\int_{0}^{z} \frac{d z^{\prime}}{H\left(z^{\prime}\right)}\right)^{2} \frac{z}{H(z)}\right]^{1 / 3}
\end{equation}

Since the release of the seven-year WMAP data, the acoustic scale has been precisely measured by the SDSS-\uppercase\expandafter{\romannumeral 3} Baryon Oscillation Spectroscopic Survey (BOSS) galaxy surveys and Sloan Digital Sky Survey (SDSS), and the WiggleZ and 6dFGRS survey. The measurement results summarized from the previous work are shown in Table \ref{tab02}.
\begin{table}[h]
    \centering
    \caption{BAO data in the Nine-Year analysis}\label{tab02}
    \begin{tabular}{cccc}
\toprule
Red shift $z$ & Data Set & $r_{\mathrm{s}}\left(z_{d}\right) / D_{v}(z)$ & Reference \\

\midrule

0.1 & 6dFGRS & $0.336 \pm 0.015$ & Beutler et al. (2011) \cite{bib55} \\
0.35 & SDSS-DR7-rec & $0.113 \pm 0.002$ & Padmanabhan et al. (2012) \cite{bib56} \\
0.57 & SDSS-DR9-rec & $0.073 \pm 0.001$ & Anderson et al. (2012) \cite{bib57} \\
0.44 & WiggleZ & $0.0916 \pm 0.0071$ & Blake et al. (2012) \cite{bib58} \\
0.60 & WiggleZ & $0.0726 \pm 0.0034$ & Blake et al. (2012) \cite{bib58} \\
0.73 & WiggleZ & $0.0592 \pm 0.0032$ & Blake et al. (2012) \cite{bib58} \\

\bottomrule
    \end{tabular}
\end{table}

\begin{figure}
    \centering
    \includegraphics[width=0.6\linewidth]{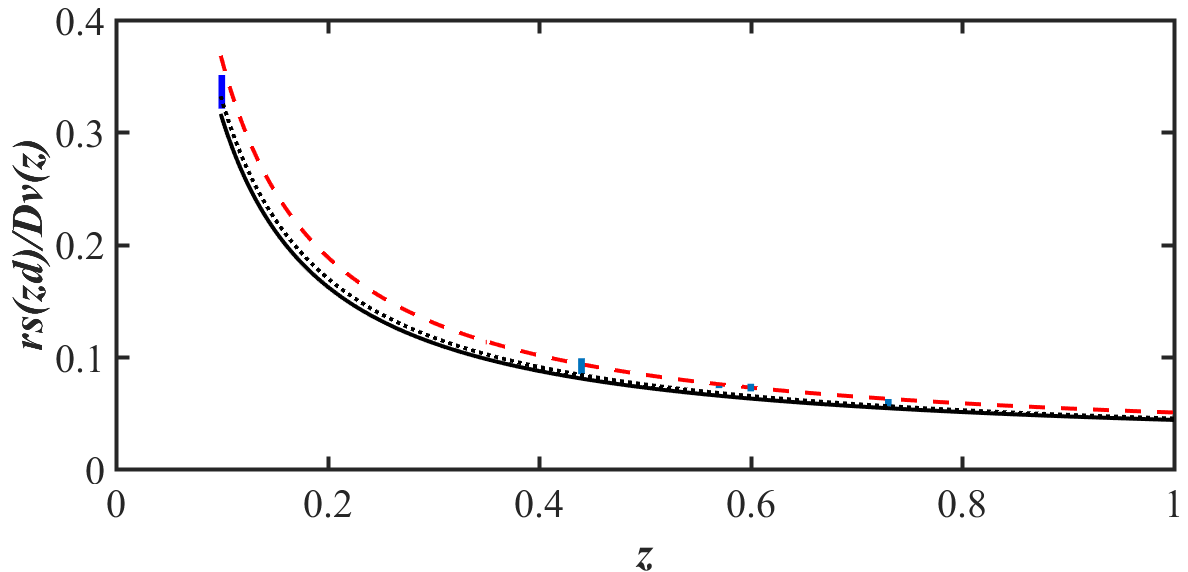}
    \caption{The comparison of the $\mathrm{z} \sim r_{\mathrm{s}}\left(z_{d}\right) / D_{v}(z)$ curves. The solid line corresponds to the $\Lambda \mathrm{CDM}$ model with $\Omega_{\mathrm{m} 0}=0.295, \Omega_{\Lambda 0}=0.705$, the dotted line corresponds to the $\Lambda C D M$ model with the best estimated parameters (i.e., $\Omega_{\mathrm{m} 0}=\mathbf{0 . 2 6 6}, \boldsymbol{\Omega}_{\boldsymbol{\Lambda} 0}=\mathbf{0 . 7 3 4}$ ) obtained from the BAO data alone, the dashed line corresponds to the bulk viscosity model with $\boldsymbol{\Omega}_{\mathrm{m} 0}=\mathbf{0 . 2 9 5}, \boldsymbol{\Omega}_{\boldsymbol{\Lambda} 0}=\mathbf{0 . 7 0 5}$, $\Omega_{\tau}=\mathbf{0 . 0 1 8}$. The observational data $r_{\mathrm{s}}\left(z_{d}\right) / D_{\mathrm{v}}(z)$ are also plotted with their error bars.}
    \label{fig03}
\end{figure}

Similarly, here we used the parameters obtained from the SNe Ia data to study the BAO data. Figure \ref{fig03} shows the comparison of the $\mathrm{z} \sim r_{\mathrm{s}}\left(z_{d}\right) / D_{v}(z)$ curves derived from the two models.

In addition, the obtained value of $\chi^{2}$ corresponding to the $\Lambda \mathrm{CDM}$ model (the solid line in Figure \ref{fig03}) is 131.87 , while for the dotted line, the obtained value of $\chi^{2}$ is 65.10 , and for the bulk viscosity model (the dashed line), the obtained value of $\chi^{2}$ is 11.33 , which imply that the bulk viscosity model has a significantly better fitting to the combined SNe Ia and BAO data than the $\Lambda \mathrm{CDM}$ model.

\subsection{Data from Hubble parameter}

Based on different ages of galaxies, we can obtain the observational Hubble parameter (see for instance Ref. \cite{bib59}). The function of $H(z)$ plays an important role in understanding the evolution of universe and the property of dark energy, since its value is directly obtained from the cosmological observations. So the Hubble parameter can be applied to constrain the cosmological parameters independently. Here we take the data coming from Ref. \cite{bib60,bib61,bib62,bib63,bib64,bib65,bib66,bib67,bib68,bib69,bib70,bib71,bib72,bib73,bib74}, which are shown in Table \ref{tab03}, to study the two models.

The $\chi^{2}$ function of the Hubble parameter is defined by
\begin{equation}
    \label{eq25}
\chi_{H}^{2}=\sum_{i=1}^{41} \frac{\left[H\left(z_{i}\right)_{o b s}-H\left(z_{i}\right)_{t h}\right]^{2}}{\sigma^{2}\left(z_{i}\right)}
\end{equation}

And the corresponding goodness-of-fit $\chi^{2}$ dof is defined as
\begin{equation}
    \label{eq26}
\chi_{d o f}^{2}=\chi^{2} /(N-n)
\end{equation}

Where $N$ denotes the number of the observed data and $n$ is the number of parameters involved in the applied model. And $\chi^{2}_{dof}$  $\leq 1$ represents the fit is good and the data are consistent with the applied model.

\begin{figure}[H]
    \centering
    \includegraphics[width=0.6\linewidth]{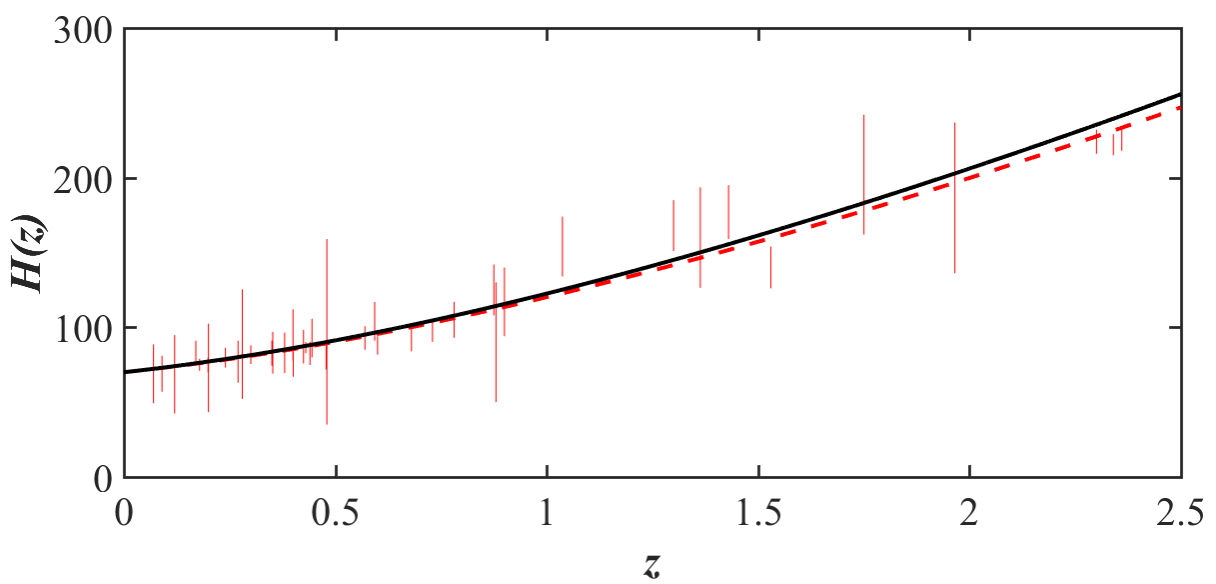}
    \caption{The comparison of the $\mathrm{z} \sim H(z)$ curves. The solid line corresponds to the $A C D M$ model with $\Omega_{\mathrm{m} 0}=0.295, \Omega_{\Lambda 0}=0.705, H_{0}=70 \mathrm{~km} / \mathrm{s} / \mathrm{Mpc}$, the dashed line corresponds to the bulk viscosity model with $\Omega_{\mathrm{m} 0}=0.295, \Omega_{\Lambda 0}=0.705, \Omega_{\tau}=0.018, H_{0}=70 \mathrm{~km} / \mathrm{s} / \mathrm{Mpc}$. The $H_{\mathrm{obs}}(z)$ data are also plotted with their error bars.}
    \label{fig04}
\end{figure}

Now, with the above best estimated parameters obtained from the SNe Ia data alone, i.e., in the $\Lambda \mathrm{CDM}$ model $\Omega_{\mathrm{m} 0}=0.295, \Omega_{\Lambda 0}=0.705, H_{0}=70 \mathrm{~km} / \mathrm{s} / \mathrm{Mpc}$ and in the bulk viscosity model $\Omega_{\mathrm{m} 0}=0.295, \Omega_{\Lambda 0}=0.705, \Omega_{\tau}=0.018$ (within the range of $\Omega_{\tau}=0.01 \pm 0.01$ ), $H_{0}=70 \mathrm{~km} / \mathrm{s} / \mathrm{Mpc}$, we would like to compare the two models. The comparison of the $\mathrm{z} \sim H(z)$ curves derived from the two models are shown in Figure \ref{fig04}.

Correspondingly, for the $\Lambda$ CDM model, $\chi^{2}=30.81, \chi^{2}_{ dof}=30.81 /(41-3)=0.81<1$. While for the bulk viscosity model, $\chi^{2}=20.80, \chi^{2}_{ dof}=20.80 /(41-4)=0.56<1$. Note that in Ref. \cite{bib65}, it present that with the above $41 H(z)$ data alone the obtained best estimated parameters involved in the $\Lambda \mathrm{CDM}$ model is $\Omega_{\mathrm{m} 0}=0.25, \Omega_{\Lambda 0}=0.75, H_{0}=70.6 \mathrm{~km} / \mathrm{s} / \mathrm{Mpc}$, and correspondingly, $\chi_{\min }^{2}=18.55$, which is very close to the value of $\chi^{2}$ obtained from the bulk viscosity model with the best estimated parameters obtained from the SNe Ia data. It imply that the bulk viscosity model has a better fitting to the combined SNe Ia and $41 H(z)$ data than the $\Lambda \mathrm{CDM}$ model.

\begin{table}[H]
    \centering
    \caption{The observational Hubble parameter data (in units [ $\mathrm{km} / \mathrm{s} / \mathrm{Mpc}]$ )}\label{tab03}
    \begin{tabular}{cccccccc}
\toprule

$z$ & $H(z)$ & $\sigma_{H}$ & Ref. & $z$ & $H(z)$ & $\sigma_{H}$ & Ref. \\

\midrule

0.070 & 69 & 19.6 & \cite{bib60} & 0.480 & 97 & 62 & \cite{bib69} \\
0.090 & 69 & 12 & \cite{bib61} & 0.570 & 92.9 & 7.855 & \cite{bib70} \\
0.120 & 68.6 & 26.2 & \cite{bib60} & 0.593 & 104 & 13 & \cite{bib62} \\
0.170 & 83 & 8 & \cite{bib61} & 0.6 & 87.9 & 6.1 & \cite{bib68} \\
0.179 & 75 & 4 & \cite{bib62} & 0.68 & 92 & 8 & \cite{bib62} \\
0.199 & 75 & 5 & \cite{bib62} & 0.73 & 97.3 & 7.0 & \cite{bib68} \\
0.200 & 72.9 & 29.6 & \cite{bib60} & 0.781 & 105 & 12 & \cite{bib62} \\
0.240 & 79.69 & 6.65 & \cite{bib63} & 0.875 & 125 & 17 & \cite{bib62} \\
0.270 & 77 & 14 & \cite{bib60} & 0.88 & 90 & 40 & \cite{bib69} \\
0.280 & 88.8 & 36.6 & \cite{bib61} & 0.9 & 117 & 23 & \cite{bib61} \\
0.300 & 81.7 & 6.22 & \cite{bib64} & 1.037 & 154 & 20 & \cite{bib62} \\
0.350 & 82.7 & 8.4 & \cite{bib65} & 1.300 & 168 & 17 & \cite{bib61} \\
0.352 & 83 & 14 & \cite{bib62} & 1.363 & 160 & 33.6 & \cite{bib71} \\
0.3802 & 83 & 13.5 & \cite{bib66} & 1.43 & 177 & 18 & \cite{bib61} \\
0.400 & 95 & 17 & \cite{bib60} & 1.53 & 140 & 14 & \cite{bib61} \\
0.4004 & 77 & 10.02 & \cite{bib66} & 1.75 & 202 & 40 & \cite{bib61} \\
0.4247 & 87.1 & 11.2 & \cite{bib66} & 1.965 & 186.5 & 50.4 & \cite{bib71} \\
0.43 & 86.45 & 3.68 & \cite{bib67} & 2.300 & 224 & 8 & \cite{bib72} \\
0.440 & 82.6 & 7.8 & \cite{bib68} & 2.34 & 222 & 7 & \cite{bib73} \\
0.44497 & 92.8 & 12.9 & \cite{bib66} & 2.36 & 226 & 8 & \cite{bib74} \\
0.4783 & 80.9 & 9 & \cite{bib66} &  &  &  &  \\
\bottomrule
    \end{tabular}
\end{table}

\section{Statefinder analysis}\label{sec4}
In this section, the statefinder diagnostic pair $\{r, s\}$ parameters, which is introduced by Sahni et al. \cite{bib75}, will be appiled to study the viscous universe with the cosmological constant. The statefinder is a geometrical diagnostic tool that can characterize the properties of dark energy in a model-independent manner. The statefinder parameters $\{r, s\}$ are usually defined by
\begin{equation}
    \label{eq27}    
r=\frac{\dddot{a}}{a H^{3}}, \quad s=\frac{r-1}{3\left(q-\frac{1}{2}\right)}
\end{equation}

Where $q$ denotes the deceleration parameter, which is defined as
\begin{equation}
    \label{eq28}
q(z) \equiv-\frac{\ddot{a}}{a} \frac{1}{H^{2}}
\end{equation}

\begin{figure}[H]
    \centering
    \includegraphics[width=0.6\linewidth]{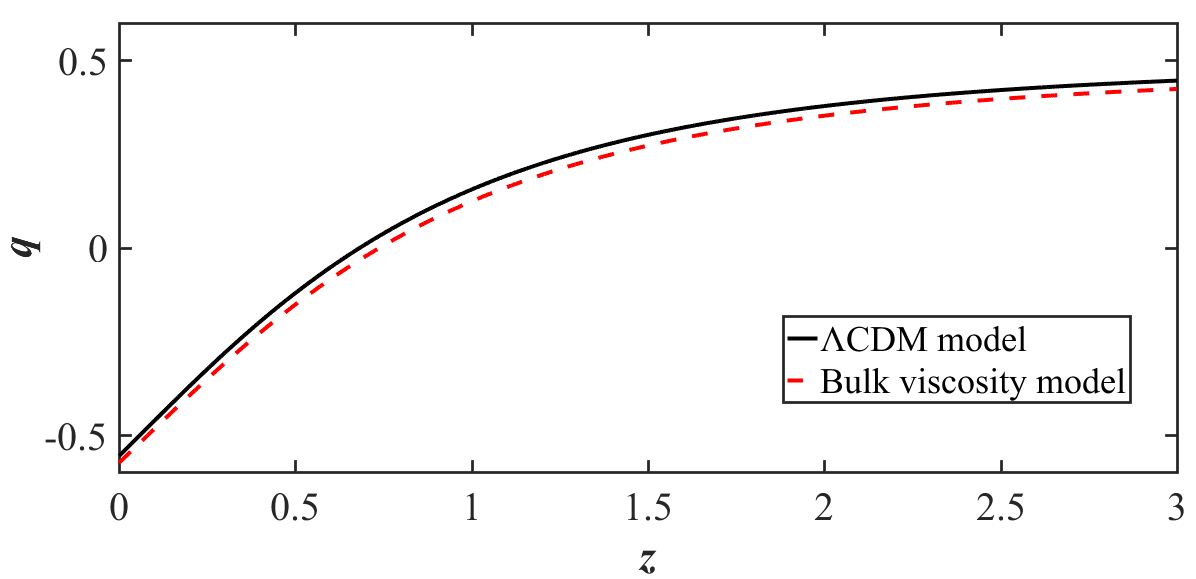}
    \caption{The comparison of the $\mathrm{z} \sim q$ curves. The solid line corresponds to the $\Lambda C D M$ model with $\Omega_{\mathrm{m} 0}=0.295, \Omega_{\Lambda 0}=0.705$, the dashed line corresponds to the bulk viscosity model with $\Omega_{\mathrm{m} 0}=\mathbf{0 . 2 9 5}, \Omega_{\Lambda 0}=0.705, \Omega_{\tau}=0.018$.}
    \label{fig05}
\end{figure}

And the $z \sim q$ curves derived from the two models with the best estimated parameters obtained from the SNe Ia are shown in Figure \ref{fig05}.

\begin{figure}[H]
    \centering
    \includegraphics[width=0.6\linewidth]{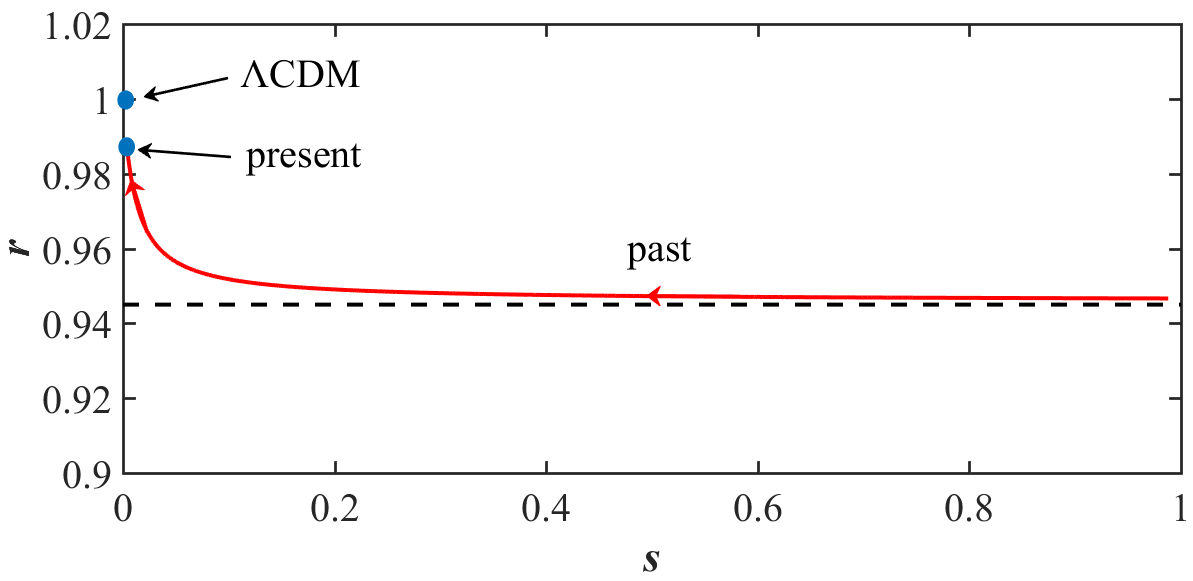}
    \caption{The evolution of the bulk viscosity model in the $r$-s plane for the best estimates of the parameters from the SNe Ia data, i.e., $\Omega_{\mathrm{m} 0}=0.295, \Omega \Lambda 0=0.705, \Omega \tau=0.018$. In the $r$-s plane the $\Lambda C D M$ model corresponds to the point $\{0,1\}$. The arrow denotes the time direction of the evolution. The dashed line is the asymptote line of the solid line.}
    \label{fig06}
\end{figure}

From Figure \ref{fig05} we can obtain that $q_{0}=-0.557, z_{T}=0.68$ for the $\Lambda \mathrm{CDM}$ model, and $q_{0}=-0.575$, $z_{T}=0.74$ for the bulk viscosity model, where $q_{0}$ is the present deceleration parameter and $z_{T}$ is the transition red shift, at which $q$ is equal to 0 . And the latest observations indicates that $q_{0}=-$ $0.63 \pm 0.12$ \cite{bib76,bib77}, which imply that the bulk viscosity model cannot be discriminated from the $\Lambda \mathrm{CDM}$ model by the deceleration parameter due to the insufficient observation accuracy.

Based on Eq. (\ref{eq27}), we can obtain the $\{r, s\}$ plane trajectories of the bulk viscosity model, which is shown in Figure \ref{fig06}.

It can be seen from Fig. 6 that the $r$-s evolution starts from a region $r \sim 0.945, s \sim 1$ and tends to the $\Lambda \mathrm{CDM}$ point in the future, and the present position of the universe corresponds to $\left\{r_{0, s_{0}}\right\}=$ $\{0.984,0.0047\}$, which is close to the $\Lambda$ CDM point. Fig. 6 shows that the bulk viscosity model is obviously different from the $\Lambda \mathrm{CDM}$.

\section{Conclusions}\label{sec5}
In this paper, we performed a detailed study of a bulk viscous matter universe with cosmological constant $\Lambda$. The explicit form of the bulk viscosity is still unknown, but we found that if we follow the assumption that $\zeta=\zeta_{1} H$, where $H$ is the Hubble parameter, $\zeta_{1}$ is a constant, then in an expanding universe, the (relative) matter density will be always greater than $8 \pi G \zeta_{1}$, i.e., 0.012 (the value is obtained from Eq. (\ref{eq13}) and the obtained $\Omega_{\tau}=0.018$ ), and in the future the (relative) matter density tends to 0.012 . The universe in the bulk viscosity model also tend to the de Sitter time-space in the future, but the expansion rate of the future's universe predicted by the bulk viscosity model is lower than which predicted by the $\Lambda \mathrm{CDM}$ model.

Secondly, with the JLA SNe Ia sample we estimated the values of parameters involved in the bulk viscosity model. And subsequently, we applied the best estimated parameters obtained from the SNe Ia data to study the CMB data, BAO data and Hubble data. Note that in this paper we used a fixed values of $\Omega_{\mathrm{m} 0}, \Omega_{\Lambda 0}, \Omega_{\tau}$ (i.e., $\Omega_{\mathrm{m} 0}=0.295, \Omega_{\Lambda 0}=0.705, \Omega_{\tau}=0.018$ ), to study the SNe Ia, $\mathrm{CMB}, \mathrm{BAO}$ and Hubble data with the purpose to directly compare the two models, and we found that the bulk viscosity model has a significantly better fitting to these data than the $\Lambda \mathrm{CDM}$ model, which indicates that the bulk viscosity model is a competitive model to fit the combined $\mathrm{SNe}$ Ia + $\mathrm{CMB}+\mathrm{BAO}+H(z)$ data.

In addition, it should be noted that the obtained best estimated value of $\zeta$ is positive, which is in consistent with the requirement of the second law of thermodynamics \cite{bib17,bib18}.

On the other hand, it shows that there is no significant difference between the curves of $z \sim q$ predicted by the two models with the same parameters. And the bulk viscosity model with the best estimated parameters obtained from the SNe Ia data predict the present deceleration parameter $q_{0}=$ -0.575 , which is within the range of the observational results, i.e., $-0.64 \pm 0.12$.

Since the universe predicted by the bulk viscosity model is similar to the standard forms of dark energy, we have analyzed the model with the statefinder parameters to distinguish it from the $\Lambda \mathrm{CDM}$ model. It shows that the evolution of the $\{r, s\}$ parameters behaves in such a way that $0<s<1,0.945<r<1$. And the present position of the bulk viscosity model in the $r$-s plane corresponds to $\left\{s_{0}, r_{0}\right\}=\{0.0047,0.984\}$, which is close to the $\Lambda \mathrm{CDM}$ point $\{0,1\}$. Hence the model is obviously different from the $\Lambda \mathrm{CDM}$ model and the other dark energy model.

\O

Here it should be noted that, in this paper we have not listed the calculation results of other cosmological parameters based on the bulk viscosity model, such as the age of the universe (the
age predicted by the bulk viscosity model is slightly higher than the age predicted by the $\Lambda \mathrm{CDM}$ model), the curvature scalar, et al., for that the bulk viscosity coefficient $\zeta$ obtained from the $\mathrm{SNe}$ Ia data is so small that the evolution or values of these cosmological parameters do not deviate significantly from which are obtained from the $\Lambda \mathrm{CDM}$ model (when $\zeta=0$ it returns to the $\Lambda \mathrm{CDM}$ model), which is exactly what we expected, as that the $\Lambda \mathrm{CDM}$ model has already achieved great success in explaining some astronomical phenomenon. And in this paper we just made a little improvement to the $\Lambda \mathrm{CDM}$ model.

\bibliographystyle{elsarticle-num-names}
\bibliography{cas-refs}

\end{document}